\documentclass[twocolumn,showpacs,preprintnumbers,amsmath,amssymb]{revtex4}
\usepackage{graphicx}% Include figure files
\usepackage{dcolumn}% Align table columns on decimal point
\usepackage{bm}% bold math

\begin{document}

\title{Strange Nonchaotic Self-Oscillator}
\author{Alexey Yu. Jalnine$^{1}$, Sergey P. Kuznetsov$^{1,2}$
        %\footnote{Electronic address: Jalnine@rambler.ru}
}
\affiliation{$^{1}$Saratov Branch of Kotelnikov's Institute of Radio-Engineering and Electronics of RAS, \\ Zelenaya 38, Saratov, 410019, Russia \\
$^{2}$Institute of Mathematics Information Technologies and Physics Udmurt State University, \\ Universitetskaya 1, Izhevsk, 426034, Russia}

\date{\today}

\begin{abstract}
An example of strange nonchaotic attractor (SNA) is discussed in a dissipative system of mechanical nature driven by constant torque applied to one of the elements of the construction. So the external force is not oscillatory, and the system is autonomous. Components of the motion with incommensurable frequencies emerge due to the irrational ratio of sizes of the involved rotating elements. We regard the phenomenon as strange nonchaotic self-oscillations, and its existence sheds new light on the question of feasibility of SNA in autonomous systems.
\end{abstract}

% PACS, the Physics and Astronomy
                             % Classification Scheme.
%\keywords{Suggested keywords}%Use showkeys class option if keyword
                              %display desired
\maketitle

The self-oscillations are commonly understood as sustained oscillatory behaviors in nonlinear dissipative systems with feedback, which are maintained due to a stationary (non-oscillatory) energy source \cite{AVK,R,J}. Thus, characteristics of the oscillations (their form, amplitude and frequency) are determined by the system itself and do not depend on the specific initial conditions (at least in some range of their variations). It is well known that the images of periodic self-oscillations are attractive limit cycles in phase space. In nonlinear dynamics and chaos theory attractors of other types are considered too, e.g. tori corresponding to sustained quasi-periodic oscillations, and strange attractors associated with chaotic self-oscillations.

Note that non-trivial attractors may not necessarily correspond to self-oscillations. For example, dynamical behaviors in nonlinear systems with periodic (or more complex) external driving are interpreted usually as forced oscillations rather than the self-oscillations, although they are associated too with attractors in the extended phase space (that is the state space supplemented with the time axis). A notable remarkable creature among them is an object called strange nonchaotic attractor (SNA), which may be regarded as somewhat intermediate between order and chaos. The epithet ``strange'' opposes SNA to the torus-attractor, a smooth object in the phase space formed by the trajectories characterized by the ergodic property. The term ``nonchaotic'' opposes SNA to the strange chaotic attractor as it does not manifest exponential sensitivity of trajectories in respect to infinitesimal perturbations, and has no positive Lyapunov exponents.

SNAs were introduced since 1984 \cite{GOPY}, and studied quite widely in relation to nonlinear systems with quasi-periodic driving (for example, driving with combination of two or more signals with irrational ratios of the basic frequencies) \cite{FKP}. However, attempts to observe SNAs in autonomous systems, where components with incommensurable frequencies would arise not from the external driving but generated in the system in a natural way were unsuccessful \cite{AVS,Pik,Jap}. Apparently, the consensus is that the SNAs, as typical objects, do not occur in autonomous dynamical systems.

The concept of strange nonchaotic self-oscillations, which we intend to discuss in this article, corresponds to a somewhat different and more physical aspect of the problem; indeed, as noted, the concepts of attractors and self-oscillations are not identical. We will deal here with a class of systems of mechanical nature, in which the incommensurable frequencies may appear due to an irrational ratio of sizes of the rotating elements involved in the motion, while the external driving is not oscillatory being implemented by the applied torque, which is constant in time. Such systems may be represented by pendulums interacting via the belt or friction transmissions between the rotating shafts or disks attached to them, or by vehicles equipped with wheels of different sizes performing motions on a rough surface without slip.

\begin{figure}[htbp]
\includegraphics[width=3in]{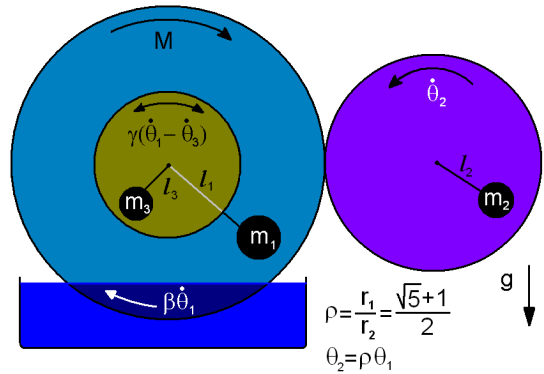}
\caption{(Color online) Schematic representation of a mechanical system able to manifest strange nonchaotic self-oscillations}
\label{fig:f1}
\end{figure}

For a simple system of dissipative pendulums with frictional transmission and constant torque driving we will demonstrate a sustained dynamical behavior, which must be regarded as self-oscillatory according to the basic definition, being associated with SNA in the phase space of the system described by ordinary differential equations.

Consider a set of disks {\bf 1}, {\bf 2}, {\bf 3} mounted in a vertical plane (Fig.~\ref{fig:f1}); two of them ({\bf 1} and {\bf 3}) are coaxial and undergo mutual viscous friction proportional to the relative angular velocity. The motion is provided by constant, not varying in time, torque applied to the disk {\bf 1}, which touches the disk {\bf 2}, so that frictional transmission of rotation without slipping takes place. In addition, the disk {\bf 1} undergoes viscous friction in respect to the rotation proportional to its angular velocity. For simplicity, we assume that the inertial properties of the system are completely provided by the point masses $m_1$, $m_2$, $m_3$ attached to the disks on the distances $l_1$, $l_2$, $l_3$ from the respective disk axes.

In essence, this is a system of pendulums with imposed mechanical constraint. The ratio of radii of the disks {\bf 1} and {\bf 2} connected through the friction transmission is supposed to be defined by an irrational number $\rho=r_1/r_2$. The condition of motion without slip of the rotating disks is expressed by the relation to the angular coordinates $\theta_2=\rho\theta_1+u$ and to the angular velocities $\dot{\theta}_2=\rho\dot{\theta}_1$. Taking this into account, we can write down the Lagrange function of the system as dependent only on the angular coordinates $\theta_{1,3}$ and velocities $\dot{\theta}_{1,3}$:
\begin{equation}
\begin{array}{ll}
L=\frac{1}{2}m_1 l_1^2 \dot{\theta}_1^2+\frac{1}{2}m_2 l_2^2 \rho^2 \dot{\theta}_1^2+\frac{1}{2}m_3 l_3^2 \dot{\theta}_3^2+m_1 l_1 g \cos{\theta_1} \\
+m_2 l_2 g \cos{(\rho \theta_1+u)}+m_3 l_3 g \cos{\theta_3}.
\end{array}
\label{eq:eq1}
\end{equation}
Introducing dissipation via the Rayleigh function:
\begin{equation}
\begin{array}{lll}
R=\frac{1}{2}\gamma_0 (\dot{\theta}_1-\dot{\theta}_3)^2+\frac{1}{2}\beta_0\dot{\theta}_1^2 -M_0\dot{\theta}_1,
\end{array}
\label{eq:eq2}
\end{equation}
we obtain the equations of motion on the form \cite{GPS}
\begin{equation}
\frac{d}{dt}\left(\frac{\partial{L}}{\partial{\dot{\theta}}_i}\right)= \frac{\partial{L}}{\partial{\theta_i}}-\frac{\partial{R}}{\partial{\dot{\theta}_i}}, i=1,3,
\label{eq:eq3}
\end{equation}
or
\begin{equation}
\begin{array}{lll}
(m_1 l_1^2+m_2 l_2^2 \rho^2) \ddot{\theta}_1=-m_1 l_1 g \sin{\theta_1} \\
-m_2 l_2 g \rho \sin{(\rho \theta_1+u)}+\gamma_0 \dot{\theta}_3-(\beta_0 +\gamma_0) \dot{\theta}_1+M_0, \\
m_3 l_3^2 \ddot{\theta}_3=-m_3 l_3 g \sin{\theta_3}+\gamma_0 (\dot{\theta}_1-\dot{\theta}_3).
\end{array}
\label{eq:eq4}
\end{equation}
Using normalized time $\tau=t\sqrt{\frac{m_1 l_1 g}{m_1 l_1^2+m_2 l_2^2 \rho^2}}$ and dimensionless parameters
\begin{displaymath}
\begin{array}{lll}
\gamma=\frac{\gamma_0}{\sqrt{m_1 l_1 g(m_1 l_1^2+m_2 l_2^2 \rho^2)}}, \beta=\frac{\beta_0}{\sqrt{m_1 l_1 g(m_1 l_1^2+m_2 l_2^2 \rho^2)}}, \\
\lambda_2=\frac{m_2 l_2}{m_1 l_1 \rho}, \lambda_3=\frac{m_3 l_3}{m_1 l_1 \rho},
\mu=\frac{m_3 l_3^2}{m_1 l_1^2+m_2 l_2^2 \rho^2}, M=\frac{M_0}{m_1 l_1 g},
\end{array}
\end{displaymath}
we obtain the equations:
\begin{equation}
\begin{array}{lll}
\ddot{\theta}=-\sin{\theta}-\lambda_2 \sin{(\rho \theta+u)}+\gamma \dot{\varphi}-(\beta+\gamma) \dot{\theta}+M, \\
\mu \ddot{\varphi}=-\lambda_3 \sin{\varphi}+\gamma_0 (\dot{\theta}-\dot{\varphi}),
\end{array}
\label{eq:eq5}
\end{equation}
where $\theta=\theta_1$, $\varphi=\theta_3$. Under the condition $\mu \ll 1$ the equations are reduced to
\begin{equation}
\begin{array}{lll}
\dot{\theta}=\omega, \\
\dot{\omega}=-\sin{\theta}-\lambda_2 \sin{(\rho \theta+u)}-\lambda_3\sin{\varphi}-\beta\omega+M, \\
\dot{\varphi}=-\lambda_3 \gamma^{-1}\sin{\varphi}+\omega.
\end{array}
\label{eq:eq6}
\end{equation}

In what follows we will investigate the model~(\ref{eq:eq6}) fixing $\lambda_3=1$, $\beta=1$, $\gamma=1$, $\rho=(\sqrt{5}+1)/2$, and varying the parameters $\lambda_2$ and $M$.

\begin{figure}[htbp]
\includegraphics[width=3.3in]{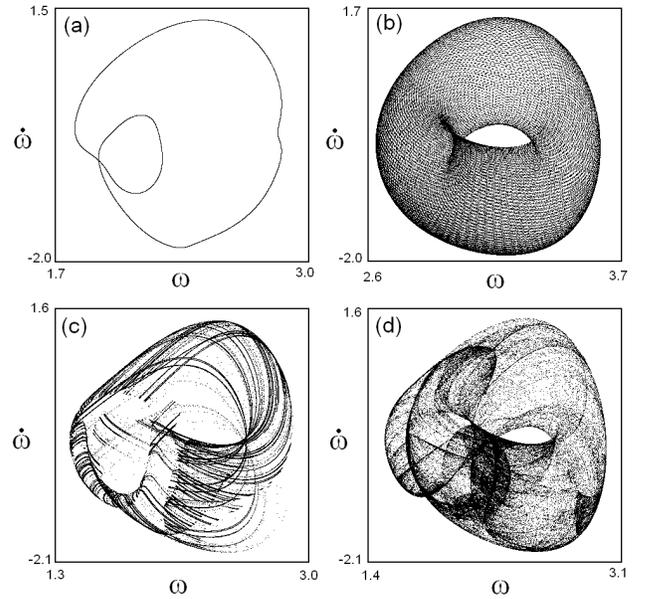}
\caption{Attractors in projection onto the plane in the Poincar\'{e} section $\theta\pmod{2\pi}=0$ for $\lambda_2=0.8$ corresponding to a two-dimensional torus at $M=2.3$ (a), a three-dimensional torus at $M=3.0$ (b), a strange nonchaotic attractor at $M=2.1$ (ñ), and chaotic attractor at $M=2.2$ (d).}
\label{fig:f2}
\end{figure}

Fig.~\ref{fig:f2} shows examples of attractors of the system~(\ref{eq:eq6}) 
depicted as projections of the cross-sections of the attractors at instants 
when the phase variable $\theta_n=\theta_0+2\pi n, n=1 \ldots 10^6$. 
In panel (a) one can see a smooth closed invariant curve, which corresponds 
to a cross-section of the attractor being a two-frequency torus. Panel (b) 
corresponds to a three-frequency torus-attractor; its section gives rise 
to a smooth two-dimensional toral surface. Attractors in panels (c) and 
(d) are strange, and for their identification the dynamic and metric 
characteristics have to be evaluated (Lyapunov exponents, 
phase sensitivity, fractal dimensions). 
As we will see, the first of them is SNA, and the other is 
a chaotic attractor.

Calculation of Lyapunov exponents was carried out in accordance with the well-known algorithm \cite{BGGS}, for which the system~(\ref{eq:eq6}) was linearized:
\begin{equation}
\begin{array}{lll}
\dot{\tilde{\theta}}=\tilde{\omega}, \\
\dot{\tilde{\omega}}=-\tilde{\theta}\cos{\theta}-\lambda_2(\rho\tilde{\theta}+\tilde{u})\cos{(\rho \theta+u)}-\lambda_3\tilde{\varphi}\cos{\varphi}-\beta\tilde{\omega}, \\
\dot{\tilde{\varphi}}=-\lambda_3 \gamma^{-1}\tilde{\varphi}\cos{\varphi}+\tilde{\omega}.
\end{array}
\label{eq:eq7}
\end{equation}
Next, together with the system~(\ref{eq:eq6}), a set of three copies of the variation equations~(\ref{eq:eq7}) with the vectors $\{\tilde{\theta}^{(k)},\tilde{\omega}^{(k)},\tilde{\varphi}^{(k)}\}_{k=1,\ldots,3}$ and $\tilde{u}^{(k)}=0$ were integrated numerically, subjected to the procedure of Gram-Schmidt orthogonalization and normalization at successive steps of the integration. The logarithms of the normalizing coefficients were summed and averaged coefficients resulting in a set of three Lyapunov exponents.

For the two-frequency torus in Fig.~\ref{fig:f2}(a) the Lyapunov exponents are  $\Lambda_1=0 \pm 0.000003, \Lambda_2=-0.0979, \Lambda_3=-0.798$ (there is one zero and others negative exponents). For the three-frequency torus of Fig.~\ref{fig:f2}(b) we have $\Lambda_1=0 \pm 0.000003, \Lambda_2=0 \pm 0.000003, \Lambda_3=-0.937$ (two zero exponents and a negative one).

Attractor in Fig.~\ref{fig:f2}(c) is characterized by a set of Lyapunov exponents $\Lambda_1=0 \pm 0.00001, \Lambda_2=-0.105, \Lambda_3=-0.894$, that indicates its nonchaotic nature. Finally, the chaotic attractor in Fig.~\ref{fig:f2}(d) has a positive, a zero, and a negative exponent: $\Lambda_1=0.0206, \Lambda_2=0 \pm 0.0001, \Lambda_3=-0.869$. 

\begin{figure}[htbp]
\includegraphics[width=3.3in]{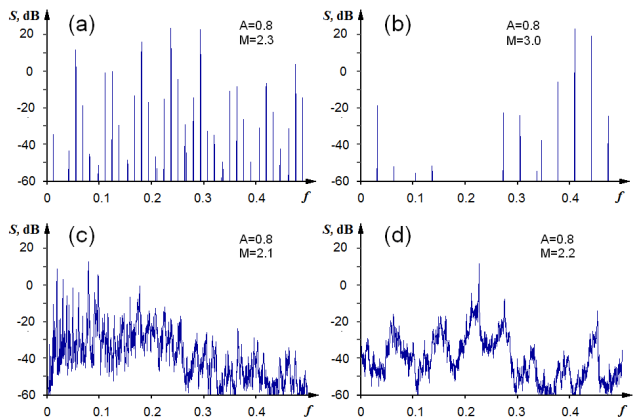}
\caption{The power spectra calculated for the variable $\dot{\theta}$ for the system~(\ref{eq:eq6}) in the case of $\lambda_2=0.8$: (a) $M=2.3$, two-dimensional torus, (b) $M=3.0$, three-dimensional torus, (c) $M=2.1$, SNA, (d) $M=2.2$, chaotic attractor.}
\label{fig:f3}
\end{figure}

Characteristic power spectra for the respective oscillation modes are shown in Fig.~\ref{fig:f3}. The spectrum is discrete for the two- and three-frequency quasi-periodic modes (panels (a) and (b)), discrete-continuous for the strange nonchaotic self-oscillations (panel (c), cf. \cite{PZFK,FKP}), and it is continuous for the chaotic regime (panel (d)).

\begin{figure}[htbp]
\includegraphics[width=3.3in]{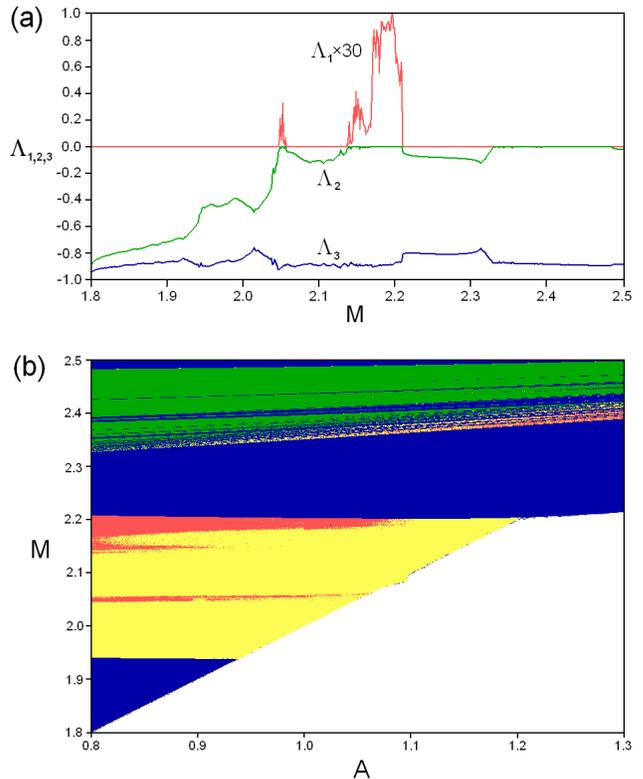}
\caption{(Color online) (a) Plots of the Lyapunov exponents 
versus parameter $M$ at $\lambda_2=0.8$. (b) chart of dynamical 
regimes for the system~(\ref{eq:eq6}), where blue areas correspond to of two-frequency tori, green designates the three-frequency tori, yellow means SNA, and red regions correspond to chaos.}
\label{fig:f4}
\end{figure}

Fig.~\ref{fig:f4}(a) depicts the Lyapunov exponents versus parameter $M$ for a fixed value of $\lambda_2=0.8$. 
This allows to reveal exactly intervals of chaotic dynamics, 
where the senior Lyapunov exponent is positive, and intervals 
of 3-torus, where two zero and one negative exponents exist. 
As well, this diagram makes it possible to guess existence of 
SNA taking into account the degree of brokenness of the parameter 
dependences for the nontrivial exponents. This brokenness appears 
as a consequence of the parametric sensitivity (structural instability) 
of SNA to variations in the control parameter of the system responsible 
for the intensity of the constant external driving.

Fig.~\ref{fig:f4}(b) gives a more detailed picture of the parameter 
space structure for the system~(\ref{eq:eq6}). There we present 
a fragment of the parameter plane chart where the ``interesting'' 
dynamics occur, including different transitions between regular 
and ``strange'' dynamic modes, and, probably critical phenomena of 
codimension 2 similar to those discussed in \cite{Kuz,KN}. 
The blue color represents the areas of two-frequency tori 
(2T), green designates the three-frequency tori (3T), 
yellow means the strange nonchaotic attractor (SNA), 
and red color corresponds to chaos (CA). In the white 
area below the line $M=\lambda_2+1$, the attractor is 
trivial stable equilibrium point.

\begin{figure}[htbp]
\includegraphics[width=3.3in]{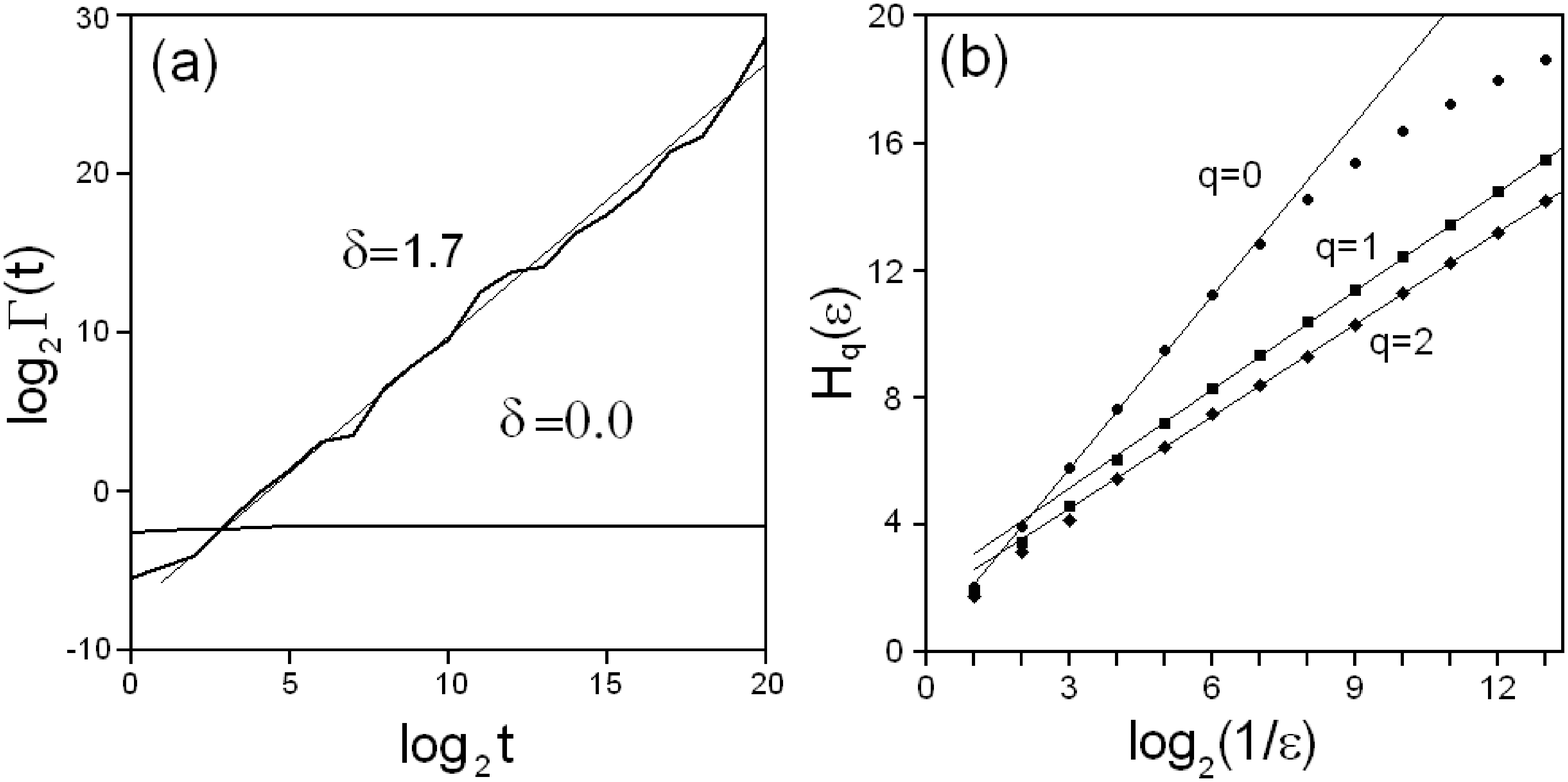}
\caption{(a) Plot of the phase sensitivity function for the SNA mode ($\delta=1.7$) and for the 2-frequency torus ($\delta=0$). (b) The dependence of Renyi entropy $H_q(\varepsilon)$ on the partition scale $\varepsilon$ for $q=0,1,2$.}
\label{fig:f5}
\end{figure}

In order to identify regions of existence of SNA with certainty, 
distinguishing them from domains of the two-frequency tori, 
which have the same signature of the Lyapunov spectrum $\{0,-,-\}$, 
we use the phase sensitivity method \cite{PF,FKP}. For this, 
in the linearized system~(\ref{eq:eq7}) we introduced the additional 
infinitesimal phase shift $\tilde{u}=const \neq 0$, and then 
the equations~(\ref{eq:eq7}) were integrated together 
with~(\ref{eq:eq6}) and with initial conditions 
$\tilde{\theta}(0)=0, \tilde{\omega}(0)=0, \tilde{\varphi}(0)=0, \tilde{u}(0)=1$. 
Now, define a piecewise smooth function as magnitude of the maximal 
variation of the variables along the orbit segment, namely, 
$\Gamma_{max}(T)=\max_{t\in[0,T]}\sqrt{\tilde{\theta}^2(t)+\tilde{\omega}^2(t)+\tilde{\varphi}^2(t)}$. 
Next, following~ \cite{PF}, we introduce the phase sensitivity function as 
minimum 
over the functions $\Gamma_{max}(T)$ computed along a set of $N$ 
trajectories with randomly specified initial conditions: 
$\Gamma(T)=\min_{(\theta_n(0),\omega_n(0),\varphi_n(0))_{n=1,\ldots,N}}\Gamma_{max}(T)$. 
It is known that the function of the phase sensitivity is bounded 
when the attractor is a smooth torus, and increases without limit 
according to a power law $\Gamma(T)\varpropto T^{\delta}$, 
where $\delta>0$ is the index of the phase sensitivity in 
the case of SNA. Typical plots of $\Gamma(T)$ 
for a smooth two-frequency torus $(\delta=0)$ and for an 
SNA $(\delta=1.7)$ are shown inFig.~\ref{fig:f5}(a). 
The parameter values are the same as in Fig.~\ref{fig:f2}(a),(c).

Direct verification of the ``strange'' geometric structure of attractor can be performed by calculating the fractal dimensions \cite{O}. The spectrum of generalized dimensions is introduced via the Renyi entropy values $H_q(\varepsilon)$ depending on the parameter $q$:
\begin{equation}
H_q(\varepsilon)=\frac{1}{1-q}\log\left(\sum_{i=1}^{N(\varepsilon)}p_i^q\right), D_q=-\lim_{\varepsilon\rightarrow 0}\frac{H_q(\varepsilon)}{\log{\varepsilon}}.
\label{eq:eq8}
\end{equation}
Here $\varepsilon$ is a size of elements covering the attractor, 
$p_i$ is the measure 
(the probability of visiting) attributed to the $i$-th element. 
With $q=0,1$, and $2$ we get the capacitance, information, 
and correlation dimension, respectively. (It should be noted 
that with $q=1$ the l'Hopital rule has to be applied in 
formulas~(\ref{eq:eq8}) to exclude the uncertainty.) 
It is believed \cite{DGO} that the dimensions for the 
strange nonchaotic attractor are $D_0=2, D_1=1$ and $D_2<1$.

To calculate the dimensions we perform the Poincar\'{e} section 
for trajectories on the attractor at $\theta_n=\theta_0+2\pi n, 
n=1,\ldots,10^7$. Next, at given $q=0,1,2$ we plot the Renyi 
entropies $H_q(\varepsilon)$ versus $\varepsilon$ and select 
linear parts of the plots there (see Fig.~\ref{fig:f5}(b)); 
the slope coefficient just yields the respective fractal 
dimension $D_q$. The following values were obtained: 
$D_0=1.8$, $D_1=1.02$, $D_2=0.96$, what reasonably 
agrees with the estimation cited above.

Thus, it is shown that the nonchaotic oscillatory regimes 
of the system~(\ref{eq:eq6}) may possess dynamic and metric 
characteristics intrinsic to SNA, and be observable in wide 
parameter ranges of the physical system of mechanical nature. 
This raises a number of issues related to the occurrence and 
destruction of SNA in the self-oscillating systems. 
In general, the ability to convert irrationally 
related spatial scales to the incommensurable 
temporal ones expands essentially the class of 
systems, which can manifest the strange nonchaotic 
dynamics. We stress that in terms of the theory of 
dynamical systems the model system~(\ref{eq:eq6}) 
formally is autonomous (with coefficients independent 
explicitly on time) in contrast to all the previously 
considered systems with SNA.

The work was supported by grant of Russian Science Foundation 
No 15-12-20035 in part of formulation and simulation of the mechanical model 
(S.P.K.) and by grant of Russian Foundation 
for Basic Research No 16-02-00135 in part of parameter space 
analysis and computations aimed to detecting and characterizing 
the SNA (A.Yu.J.).

\end{document}